\documentclass[a4paper,11pt,titlepage]{article}
\usepackage[cp1250]{inputenc}
\usepackage{amsmath,amsthm,amssymb}
\usepackage{color}
\usepackage{graphicx}

\newcommand{\OFLTS}{OF^{(LTS,n,h)}(\beta)}
\newcommand{\R}{\mathbb{R}}
\newcommand{\NN}{\mathbb{N}}
\newcommand{\XXT}{(X^{T}X)^{\!-1}}
\newcommand{\OFOLS}{OF^{(OLS,X,Y)}(\beta)}
\newcommand{\OFOLSw}{OF^{(OLS,WX,WY)}(\beta)}
\newcommand{\OLSbeta}{\hat{\beta}^{(OLS,n)}}

\newcommand{\OLSbetaW}{\hat{\beta}^{(OLS,WX,WY)}}

\newcommand{\LTSbeta}{\hat{\beta}^{(LTS,n,h)}}

\newcommand{\sumin}{\sum_{i=1}^n}
\newcommand{\sumih}{\sum_{i=1}^h}
\newcommand{\res}[1]{r_{(#1)}^2}
\newcommand{\Hset}{\mathcal{H}}
\newcommand{\Uset}{\mathcal{U}}
\newcommand{\Useq}{\mathcal{U}^{(\text{seq})}}
\newcommand{\Wmin}{W^{(\text{min})}}

\newcommand{\Q}{Q^{(n,h)}}

\newcommand{\Qp}{Q^{(n,p+1)}}

\newcommand{\n}{\in \hat{n}}
\newcommand{\m}{\in \hat{m}}

\newcommand{\D}{\mathcal{D}}

\def \proof {\emph{Proof}}
\def \proofend {\hfill \emph{Q.E.D} \vspace{0.2cm}}

\def \KKA {$\text{BSA}$}

\def \PLTS {$\text{Problem~\ref{problem:LTS}}$}

\DeclareMathOperator*{\argmin}{arg\,min}

\newtheorem{theorem}{Theorem}
\newtheorem{assertion}[theorem]{Assertion}
\newtheorem{lemma}[theorem]{Lemma}
\newtheorem{corollary}[theorem]{Corollary}
\newtheorem{assump}{Assumption}
\newtheorem{problem}{Problem}
\newtheorem{example}{Example}
\newtheorem{program}{Program}

\newtheorem{definition}{Definition}

\begin{document}
\begin{center}
    \textbf{\KKA -- exact algorithm computing LTS estimate}
\end{center}

\begin{center}
    $\text{Karel Klouda}^{1,2}$ \\[2mm]
    \texttt{karel@kloudak.eu}
\end{center}

\small{$\ ^1$ FNSPE, Czech Technical University in Prague}

\small{$\ ^2$ LIAFA, Universit\'{e} Denis-Diderot (Paris VII)}\\[0.1cm]

\begin{center}
\textbf{Abstract}
\end{center}
The main result of this paper is a new exact algorithm computing
the estimate given by the Least Trimmed Squares (LTS). The
algorithm works under very weak assumptions. To prove that, we
study the respective objective function using basic techniques of
analysis and linear algebra.

\section{Introduction}

In general, (linear) regression analysis is concerned with
problems of the following type. One random variable $Y$ called
\emph{response variable} is supposed to fit linear regression
model $Y = x^T \beta^0 + e$, where $x \in \R^p$ is a
vector\footnote{All vectors in this text are treated as column
vectors.} of \emph{explanatory variables} (random or not),
$\beta^0 \in \R^p$ is a vector of \emph{regression coefficients}
and $e$ is an \emph{error term}. The aim of regression analysis is
to estimate $\beta^0$ having $n$ measurements of $Y$ and $x$.
These measurements will be denoted as vector $Y = (Y_1, \ldots,
Y_n)$ and as a \emph{design matrix}
\begin{equation} \label{eq:maticeX}
    X = \left( \begin{array}{cccc}
        x_{1}^1 & x_{1}^2 & \ldots & x_{1}^p \\
        x_{2}^1 & x_{2}^2 & \ldots & x_{2}^p \\
        \vdots & \vdots &        & \vdots \\
        x_{n}^2 & x_{n}^2 & \ldots & x_{n}^p \\
        \end{array} \right),
\end{equation}
vector $x_i$ stands for a transposition of $i$-th row of the
matrix $X$.

The best known estimate of $\beta^0$ is the estimate given by the
(ordinary) least squares method (OLS estimate)
\begin{equation} \label{eq:LS_odhad_expl}
   \OLSbeta = \XXT X^{T}Y,
\end{equation}
which is in fact the projection of $Y$ into the linear envelope of
the columns of $X$. Unfortunately, the OLS estimate was shown to
be very sensitive with respect to data contamination of many kinds
(for more see~\cite{Rousseeuw1987}). Therefore, other estimates
which are less sensitive or, in other words, more \emph{robust}
were introduced. One of such estimates is the estimate given by
the \emph{Least Trimmed Squares} method (LTS estimate) proposed by
Rousseeuw in 1984~\cite{Rousseeuw1984}.

OLS estimate~\eqref{eq:LS_odhad_expl} is actually obtained as a
minimum of the OLS objective function (OLS-OF) defined as a sum of
squares of \emph{residuals} $r_i(\beta) = Y_i - x_i^T \beta$,
i.e., the OLS-OF reads
\begin{equation}\label{eq:OLS_OF}
    \OFOLS = \sumin (Y_i - x_i^T\beta)^2.
\end{equation}
The basic idea of the LTS method is that the contaminating data
points lay out of the main bulk of data and hence their residuals
are bigger. It means that in order to obtain a more robust
estimate of regression coefficients we ignore (trim) some portion
of data points with biggest residuals. Formally, the LTS estimate
is defined as a minimum of the LTS objective function (LTS-OF)
\begin{equation}\label{eg:OF_LTS_def}
    \OFLTS = \sumih r_{(i)}^2(\beta),
\end{equation}
where $h$ is a parameter which determines how many ($n-h$) data
points is to be trimmed and $r_{(i)}(\beta)$ stands for the $i$-th
smallest residuum at $\beta$. Since it is not reasonable to ignore
more than a half of data points, $h$ usually takes values between
$n/2$ and $n$.

\subsection{Algorithms}

As we will see in the following section, there exists a
straightforward algorithm always giving the exact value of the LTS
estimate, but it requires $n \choose h$ computations of OLS
estimates for $h$ not trimmed data points. As this algorithm (and
its modifications, see~\cite{Agullo2001}) has been the only known
exact algorithm, another faster ways how to obtain the LTS
estimates were introduced. All these faster algorithms are
probabilistic, i.e., it is not sure they return the exact value of
the LTS estimate. There exist two kinds of probabilistic
algorithms which may be described, using terminology
from~\cite{Hawkins1999}, as algorithms finding $\beta$ satisfying
the \emph{weak} and \emph{strong} necessary condition
respectively. In fact, $\beta$ satisfies the weak necessary
condition if, and only if, it is a local minimum of the LTS-OF.
Algorithms finding $\beta$'s satisfying the weak conditions have
been proposed independently several times, first such algorithm is
from~\cite{Visek1996}, its modification by the same author can be
found in~\cite{Visek2000a}, another algorithm of this type was
introduced along with the notion of weak necessary condition
in~\cite{Hawkins1999}, and a version for large data sets is
described in~\cite{Rousseeuw1999}. In the case of the strong
condition the situation is simple as there is only one
representative: Feasible Solution Algorithm \cite{Hawkins1994}.

Since we are going to study an algorithm solving the problem of
minimizing of the LTS-OF, we can forget the complex statistical
background and formulate it as follows:
\begin{problem} \label{problem:LTS}
    Find the LTS estimate
    \begin{equation} \label{eq:OF_LTS_def_problem}
        \LTSbeta = \argmin_{\beta \in \R^{p}} \sumih
        \res{i}(\beta) = \argmin_{\beta \in \R^{p}} \sumih (y_i - \beta
        x_i^T)^2,
    \end{equation}
    where $n > p \geq 1$, $Y = (y_1, \ldots, y_n)^T$, $X = (x_1, \ldots, x_n)^T$ is a matrix from
     $\R^{n,p}$, and $h$ is an integer such that $p \leq h \leq n$.

    Further, let us denote the data for which the problem is defined by
    \begin{equation*}
        \D = \{(y_i,x_i^T)\,|\,i \n\}.
    \end{equation*}
\end{problem}

Prior to introduction of the new exact algorithm, we need to study
the LTS-OF as the algorithm is based on some special properties of
it. Having described these properties, we will first propose
one-dimensional version of the algorithm which is easier to
demonstrate, then the general case will be given.

\section{Objective function}

\subsection{Discrete reformulation of LTS-OF}

For every $\beta \in \R^{p}$ only $h$ data with least squared
residuals appear in (\ref{eg:OF_LTS_def}). Every such $h$-element
subset of all data $\mathcal{D}$ can be unambiguously determined
by 0-1 vector $w \in \R^{n}$, where $w^i = 1$ if $(y_i,x_i^T)$ is
an element of this subset and $w^i = 0$ otherwise -- in this sense
we will be speaking about a \emph{subset $w$}. For any element of
the set of all such vectors
\begin{equation}\label{eq:Q_set}
    \Q = \{w \in \R^{n}|\ w^i \in \{0,1\}, i \n, w^1 + \ldots + w^n = h\},
\end{equation}
we define two sets
\begin{eqnarray}
  I_w &=& \{ k \n\,|\, w^k = 1 \}, \label{eq:I_w}\\
  \nonumber O_w &=& \{ k \n\,|\, w^k = 0 \}.
\end{eqnarray}

Clearly, for any $\beta$ there exists at least one $w \in \Q$ so
that $\sumih \res{i}(\beta) = \sumin w^i r_i^2(\beta)$. Employing
this fact we get:
\begin{eqnarray}
  \min_{\beta \in \R^{p}} \sumih r^2_{(i)}(\beta)& = & \min_{\beta \in \R^{p}, w \in \Q} \sumin w^i r^2_i (\beta) \label{eg:OF_LTS_w_to_del}\\
                & = & \min_{w \in \Q} \left( \min_{\beta \in \R^{p}} (WY - WX\beta)^T(WY - WX\beta) \right) \label{eq:OF_LTS_parabolas}\\
                & = & \min_{w \in \Q} \| WY - WX(X^T W X)^{-1}X^T WY)
                \|^2, \label{eq:OF_LTS_W}
\end{eqnarray}
where $W = \textrm{diag}(w)$. Having this equation, we can propose
a new objective function of the LTS defined on $\Q$
\begin{equation}\label{eq:JW_def}
    J(w) = \| W(Y - X(X^T W X)^{-1}X^T WY)) \|^2.
\end{equation}
It is straightforward that $J(w)$ is the minimum of the
\mbox{OLS-OF} for the subset $w$, i.e. $\min_{\beta \in
\R^{p}}OF^{(OLS,WX,WY)}(\beta)$. Finally, we can also reformulate
\eqref{eq:OF_LTS_def_problem} to the following form
\begin{equation}\label{eq:LTS_jinak_w}
    \LTSbeta = (X^T W^* X)^{-1}X^T W^* Y,
\end{equation}
where
\begin{equation} \label{eq:wstar_def}
    w^* = \argmin_{w \in \Q} J(w)\quad\mathrm{and}\quad W^* =
    \mathrm{diag}(w^*).
\end{equation}

\subsection{Domain of LTS-OF} \label{sec:domain}

The discrete version of the LTS-OF proposed in the previous
paragraph is well known and has already been described in many
articles dealing with the LTS, especially with computing the LTS
estimate. In the present paragraph we shall discuss the
non-discrete LTS-OF, i.e. $\OFLTS$, where $\beta \in \R^{p}$.
Several of the features, we are going to propose, were already
mentioned in~\cite{Visek2006a}. We will reprove them and broaden
them somewhat.

\begin{definition} \label{def:relation_Z}
We define a~relation $Z \subset \R^{p} \times \Q$ by
\begin{equation*}
    (\beta,w) \in Z \Leftrightarrow \sumih \res{i}(\beta) = \sumin
    w^i r_i^2 (\beta).
\end{equation*}
Further, we define a set $\Uset \subset \R^p$ as the set where $Z$
is a mapping from $\R^p$ to $\Q$. Complement of $\Uset$ to $\R^p$
is denoted by $\Hset$.
\end{definition}

\begin{assertion}
For $\beta \in \R^{p}$ there exists only one $w \in \Q$ so that
$(\beta,w) \in Z$, i.e., $\beta \in \Uset$, if, and only if,
$\res{h}(\beta) < \res{h+1}(\beta)$.
\end{assertion}
Indeed, if $r_i^2(\beta) = \res{h}(\beta) = \res{h+1}(\beta) =
r_j^2(\beta)$ and $(\beta,w) \in Z$ then also $(\beta, \hat{w})
\in Z$ where $\hat{w}$ is created from $w$ by swopping the $i$-th
and $j$-th elements.
\begin{corollary} \label{cor:setH}
    The following holds:
    \begin{equation*}
        \Hset = \{\beta \in \R^{p}\,|\,\res{h}(\beta) =
        \res{h+1}(\beta)\}.
    \end{equation*}
\end{corollary}
For every $\beta \in \Hset$ there exist $i,j \n$ such that
$\res{h}(\beta) = r^{2}_i(\beta) = r^2_j({\beta}) = \res{h +
1}(\beta)$, this equality is equivalent to $r_i({\beta}) = \pm
r_j({\beta}) \Leftrightarrow y_i \mp y_j + (x_i^T \mp
x_j^T){\beta} = 0$.
\begin{assump}\label{assump:hyperplane}
Let us assume that for \PLTS\ that
\begin{enumerate}
    \item $(\forall i,j \in \hat{n}, i \neq
    j)(x_i \neq \pm x_j)$,
    \item  $(\forall i \n)(||\,x_i|| \neq 0)$.
\end{enumerate}
\end{assump}
If Assumption \ref{assump:hyperplane} is fulfilled, then $y_i \mp
y_j + (x_i \mp x_j){\beta} = 0$ is represents a hyperplane, i.e.,
a closed set having Lebesgue measure~0. Since $\Hset$ is a finite
union of such sets, it is also closed and of Lebesgue measure~0.
\begin{assertion}
If Assumption \ref{assump:hyperplane} is fulfilled, we get
    \begin{enumerate}
        \item $\mu_L(\Hset) = 0$, i.e. the Lebesgue measure of
        $\Hset$ is zero,
        \item the set $\Uset$ is open.
    \end{enumerate}
\end{assertion}

Assume that for two different $\beta_1, \beta_2 \in \Uset$
$$
\{\beta \in \R^{p}\,|\, \beta = \beta_{1} + t (\beta_{2} -
\beta_{1}), t \in [0,1]\} \cap \Hset = \emptyset,
$$
i.e., the line between $\beta_{1}$ and $\beta_{2}$ does not cross
the set $\Hset$, then on this line we must have $\res{h}(\beta) <
\res{h+1}(\beta)$ and so $Z(\beta_1) = Z(\beta_2)$. In words, the
space $\R^{p}$ is ``divided'' by the set $\Hset$ into a
finite\footnote{The finiteness of the number of the subsets will
be proved later; we will prove that $m\leq {n \choose p+1}2^p$.}
number $m$ of open disjoint subsets of $\Uset$.
\begin{definition} \label{def:sequence_Ui}
    For \PLTS\ we define a sequence of $m \in \NN$ sets $\Useq =
    \{U_i\}_{i=1}^m$ such that
    \begin{enumerate}
        \item $U_i$ is open and connected\footnote{By definition, an open set $A$ is connected if it cannot be represented as the disjoint union of two or more nonempty open sets.}, for all $i = 1, \ldots, m$,
        \item $U_i \cap U_j = \emptyset$, for all $i,j, i \neq j$,
        \item $\cup_{i=1}^m U_i = \Uset$,
        \item $\cup_{i=1}^m \partial U_i = \Hset$.
    \end{enumerate}
    We say that $U_i, U_j \in \Useq$ are \emph{neighbours} if $i \neq
    j$ and $\partial U_i \cap \partial U_j \neq \emptyset$.
    Further, we define a set $\Wmin$ of $m$ vectors from $\Q$
    \begin{equation*}
        \Wmin = \{w_1,\ldots, w_m \,|\,  w_i = Z(\beta),\text{ where } \beta \in
        U_i, i \m\}.
    \end{equation*}
\end{definition}
The sequence $\Useq$ is uniquely determined by conditions 1, 2 and
3, condition~4 is implied by condition~3. The elements of $\Wmin$
are correctly defined due to the fact that $Z(\beta) =
Z(\hat{\beta})$ for all $\beta,\hat{\beta} \in U_i$ where $i \m$
arbitrary.

\subsubsection{One-dimensional example}

As the above introduced definitions and assertions are crucial for
understanding all the results of the following paragraphs, we will
demonstrate their meaning on an example. The simplest instance of
\PLTS\ is the case of $p = 1$ when the argument of the LTS-OF is a
real number $\beta \in \R^{1}$.

\begin{figure}[!ht]
    \begin{center}
        \includegraphics[width=11.51cm,height=8.15cm]{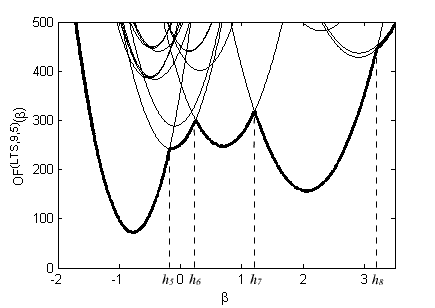}
        \caption{The bold line is the graph of the LTS-OF, the other parabolas (thin lines) are graphs of OLS-OF corresponding to various data subsets $w \in Q^{(9,5)}$.} \label{fig:LTS_OF_p1_Ui}
    \end{center}
\end{figure}

Let us assume that Assumption \ref{assump:hyperplane} is
fulfilled. Then all residuals $r^2_i = (y_i - x_i \beta)^2$, as
well as an arbitrary sum of them, are sharply convex parabolas.
Thus, for every subset $w \in \Q$ the function $\OFOLSw = \sumin
w^i(y_i - x_i \beta)^2$ is also an sharply convex parabola and the
value of the discrete function $J(w)$ is a minimum of it.
\begin{example}\label{example:p1}
    Find the LTS-estimate of \PLTS\ for the following settings:
        \begin{itemize}
            \item $n = 9$, $p = 1, h = 5$,
            \item $Y = (-0.90,-0.80,33.32,-27.23,12.63,-14.18,-3.79,-8.66,-16.45)^T$,
            \item $X =
            (1.39,-2.25,6.10,-8.50,8.26,-8.67,10.87,13.70,13.05)^T$.
        \end{itemize}
\end{example}
As $\beta$ is a scalar, it is easy to draw the graph of $\OFLTS$
for Example \ref{example:p1}. What we need is to know is how to
determine the value $\OFLTS$ for a given $\beta$. It can be easily
done by evaluating and ordering the squared residuals
$r_i^2(\beta)$ for all $i$. Employing the definition of the
relation $Z$, we can say that we need to find a subset $w$ such
that $(\beta,w) \in Z$ -- in other words, we need to find the
parabola $\OFOLSw$, corresponding to the subset $w$, such that
$\OFOLSw = \OFLTS$ for the given $\beta$. The sector of the graph
of $\OFLTS$ for Example \ref{example:p1} containing all the local
minima is depicted in Figure~\ref{fig:LTS_OF_p1_Ui}.

It is clear that for data $X$ Assumption \ref{assump:hyperplane}
is fulfilled. The second part tells us that all squared residuals
have parabolas as a graph and the first part that for any two
parabolas the intersection of their graphs is a set of Lebesque
measure~0 (a point, in the case of $p = 1$). Using our notation,
we can reformulate the last sentence in this way: the set $\Hset$
is the set of $\beta$ for which more than one parabola coincides
with the graph of the LTS-OF or, equivalently, the set of $\beta$
for which more than one subset $w$ is in the relation $Z$ with
$\beta$. Denote $\Hset = \{h_1, \ldots, h_{m-1}\}$, where $m =
\#\Useq$. For Example \ref{example:p1} $\Hset =
\{-8.16,-7.44,-6.99,-3.92,-0.18,0.25,1.21,3.20,3.84\}$.
\begin{table}[ht]
    \begin{center}
        \begin{tabular}{|c|r@{,}l|c||c|r@{,}l|c|}
        \hline
        \rule[-3mm]{0mm}{3mm}\raisebox{-1.3mm}{$i$} & \multicolumn{2}{|c|}{\raisebox{-1.3mm}{$U_i$}} & \raisebox{-1.3mm}{$\{j\in\hat{9}\,|\,w_i^j = 1\}$} & \raisebox{-1.3mm}{$i$} & \multicolumn{2}{|c|}{\raisebox{-1.3mm}{$U_i$}} & \raisebox{-1.3mm}{$\{j\in\hat{9}\,|\,w_i^j = 1\}$} \\
        \hline
        1 & $(-\infty$ & $-8.16)$ & 1,2,3,5,6 & 6 & $(-0.18$ & $0.25)$  & 1,2,5,7,8 \\
        \hline
        2 & $(-8.16$ & $-7.45)$ & 1,2,3,5,7  & 7 & $(0.25$ & $1.21)$  & 1,2,5,6,7 \\
        \hline
        3 & $(-7.45$ & $-6.99)$ & 1,2,5,6,7  & 8 & $(1.21$ & $3.20)$ &  1,2,4,5,6 \\
        \hline
        4 & $(-6,99$ & $-3.92)$ & 1,2,5,7,9 & 9 & $(3.20$ & $3.84)$ & 1,2,3,4,6 \\
        \hline
        5 & $(-3.92$ & $-0.18)$ & 1,2,7,8,9 & 10 & $(3.84$ & $+\infty)$ & 1,2,3,4,5 \\
        \hline
        \end{tabular}
    \end{center}
    \caption{The sets $U_i$ and the corresponding $w_i$ for Example~\ref{example:p1}.}\label{tab:Ui_wi_p1}
\end{table}

Regarding the set $\Uset$, we know that $\Uset = \R^{1} \setminus
\Hset$, thus the set $\Uset$ is a union of $m$ open intervals
$U_1, \ldots, U_m$. It is obvious that the sequence $\Useq$ equals
to the sequence of these intervals, i.e. $\Useq =
\{U_i\}_{i=1}^m$. All the sets $U_i$ and all the corresponding
vectors $w_i \in \Wmin$ are given in Table~\ref{tab:Ui_wi_p1}.

Note that in general we have
\begin{equation*}
    i \neq j \nRightarrow w_i \neq w_j.
\end{equation*}
For our example data $w_3 = w_7$. Note also that not all sets
$U_i$ must contain a local minimum. For us there are only 4 local
minima: (1) in $\beta = -0.77$, value 71.96 (2) in $\beta = 0.14$,
value 242.42 (3) in $\beta = 0.70$, value 246.87 and (4) in $\beta
= 2.06$, value 156.15.

\subsection{Local minima of LTS-OF} \label{sec:local_minima}

Now we will try to append to hitherto shown features and proposed
notation some others, which will be useful from the point of view
of the minimization of the objective function $\OFLTS$. Without
doubt it would be very useful to know if there exists a global
minimum or if there could be more than one local minimum. Taking
into account simultaneously the discrete form of the LTS-OF,
(\ref{eq:LTS_jinak_w}) and (\ref{eq:wstar_def}) we have the proof
of the existence of the global minimum and also the alternative
formula for it.

As for the local minima, we have to employ the notation and facts
from the previous paragraphs. We know that the domain of the
LTS-OF can be written as a union of the open sets $\Useq =
\{U_i\}^m_{i = 1}$ and the set of measure zero $\Hset$. We also
proved that for all $i = 1, \ldots, m$ and for all $\beta \in U_i$
we have $\OFLTS = OF^{(OLS,W^{i}X,W^{i}Y)}(\beta))$ where $W^{i} =
\mathrm{diag}(w_{i})$ and $w_i \in \Wmin$ (see
Definition~\ref{def:sequence_Ui}). This fact has an important
consequence.
\begin{definition}
We say that a matrix $X \in \R^{n,p}, n \geq p$, has
\emph{$h$-full rank} if a rank of the matrix $WX$ is $p$ for all
$w \in \Q$, $W = \mathrm{diag}(w)$.
\end{definition}
It is well known that the OLS estimate is unique if, and only if,
the design matrix has rank $p$. As we compute the OLS estimate for
$h$-element subset, we need the previous definition.
\begin{assertion} \label{asser:loc_minima1}
The objective function of \PLTS\ $\OFLTS$ has a local minimum in
$\beta_0 \in U_i \in \Useq$ if, and only if, the function
$OF^{(OLS,W^{i}X,W^{i}Y)}(\beta))$ has a local minimum in $\beta_0
\in U_i, i \m$.

Moreover, if $X$ has $h$-full rank, then
\begin{equation*}
    \OFLTS \text{ has a local minimum at } \beta \in U_i
    \Leftrightarrow
    \beta = \hat{\beta}^{(OLS,W^{i}X,W^{i}Y)},
\end{equation*}
where $W^i = \text{diag}(w_i)$.
\end{assertion}
How strong is the assumption  that $X$ has $h$-full rank? It
depends on the values of parameters $p$ and $h$ which determine
the dimensions of the sub-matrixes of $X$. If $h \gg p$ (usually
true), then the assumption is very weak.

Assertion~\ref{asser:loc_minima1} tells us how to find local
minima located in the open set $\Uset$. What if a local minimum is
in the set $\Hset$? In what follows, we will prove that even if a
local minimum is in the set $\Hset$, it can be still found as the
OLS estimate for some subset $w \in \Q$.
\begin{lemma} \label{lemma:convex_functions}
\begin{enumerate}
    \item Function $f:\R^{n} \rightarrow \R^{1}$, $f \in
    C^{(2)}$ is a strictly convex function if and only if
    $\nabla^2 f(x)$ is a positive-definite matrix.
    \item A strictly convex function has maximally one strict minimum.
\end{enumerate}
\end{lemma}
This Lemma is a classical result of mathematical analysis.
\begin{lemma} \label{lemma:parabolas_are_strictly_convex}
    For \PLTS\ and every subset $w \in \Q$ it holds that if the matrix $WX$ has full rank ($W = \text{diag}(w)$),
    then the function $\OFOLSw$ is strictly convex.
\end{lemma}
The proof follows from from Lemma~\ref{lemma:convex_functions} and
from the fact that $\nabla^2 \OFOLSw = X^TW X$, where $X^TW X$ is
positive-definite.
\begin{lemma} \label{lemma:obvious_on_mimima}
    Let functions $f_1, \ldots, f_k$ be continuous having
    unique strict minimum, $f_i:\R^{p} \rightarrow \R^{1}$, and let $h(x) =
    \min\{f_1(x),\ldots,f_k(x)\}$ for all $x \in \R^{p}$. Define a set
    $S = \{x \in \R^{p}\,|\,f_1(x) = \cdots = f_k(x) \}$.
    If $h$ has a local minimum at $x_0 \in S$, then $f_i$ has the
    strict global minimum at $x_0$ for all $i = 1, \ldots, k$.
\end{lemma}
\proof

    If $h$ has a local minimum at $x_0 \in S$, then there exists a
    neighbourhood $U_{(x_0)}$ of $x_0$ such that
    $$
        (\forall x \in U_{(x_0)})(h(x) \geq h(x_0) = f_1(x_0) \cdots =
        f_k(x_0)).
    $$
    The same inequality is clearly true on $U_{(x_0)}$ for all $f_i(x)$
    (note that $h(x) \leq f_i(x)$), and hence, due to the fact that all $f_i(x)$ have only one strict minimum,
    $x_0$ has to be also the point of the global
    minima of the functions $f_i, i = 1, \ldots, k$.

\proofend

Now we can propose the following not-surprising but important
theorem.
\begin{theorem} \label{asser:completeness_of_Wi}
If the matrix $X$ from \PLTS\ has $h$-full rank, then for every
local minimum at a point $\beta_0$ of the objective function
$\OFLTS$ there exists a vector $w \in \Wmin$ such that
\begin{equation*}
    \beta_0 = (X^T W X)^{-1}X^T WY,
\end{equation*}
where $W = \mathrm{diag}(w)$.
\end{theorem}

\proof

If $\beta_0 \in \Uset$, the proof follows directly from
Assertion~\ref{asser:loc_minima1}.

Let us assume that $\beta_0 \in \Hset$. It means that there exist
more than one subset being in relation $Z$ with $\beta_0$. Let us
denote these subsets by $w_{i_1}, \ldots, w_{i_k}, k \geq 2$. Now
employing the previous lemma -- putting $f_j =
OF^{(OLS,W_{i_j}X,W_{i_j}Y)}(\beta)$ for all $j$ -- and
Lemma~\ref{lemma:parabolas_are_strictly_convex} we can be sure
that $\beta_0$ is a point of global minima of all functions $f_j$.
From Definition~\ref{def:sequence_Ui} we also know that $\Hset =
\cup_{i \m} \partial U_i$. Thus, taking into account that $\beta_0
\in \Hset$, there are at least two of the subsets $w_{i_1},
\ldots, w_{i_k}, k \geq 2$ (corresponding to two neighbours from
$\Useq$ -- see Definition~\ref{def:sequence_Ui}) which are
elements of $\Wmin$.

\proofend

\section{Borders Scanning Algorithm -- \KKA}

In the present section we shall propose a new algorithm for
solving Problem~\ref{problem:LTS}. A~principle of the algorithm is
quite simple. It is based on the fact that
\begin{equation}\label{eq:OF_LTS_reformulation}
    \begin{split}
    \OFLTS  & = \min_{w \in \Q} \OFOLSw \\
            & = \min_{i =1,\ldots m}OF^{(OLS,W^iX,W^iY)}(\beta),
    \end{split}
\end{equation}
where $W = \text{diag}(w), W^i = \text{diag}(w_i)$ and $m$ and
$w_i \in \Wmin$ are defined in Definition~\ref{def:sequence_Ui}.
This equation claims that to get complete knowledge of the
complicated function \mbox{LTS-OF}, it is enough to evaluate
\emph{only} $m$ sharply convex objective functions of the OLS
estimate~$OF^{(OLS,W_iX,W_iY)}(\beta)$. Taking into
account~\eqref{eq:OF_LTS_W}, \eqref{eq:JW_def} and
Theorem~\ref{asser:completeness_of_Wi}, we can
reformulate~\eqref{eq:OF_LTS_reformulation} as
\begin{equation*}
    \min_{\beta \in \R{p}}\sumih \res{i}(\beta) = \min_{w \in \Q} J(w)
    = \min_{i \m}J(w_i),
\end{equation*}
i.e. the LTS estimate for \PLTS\ can be obtained by evaluating
$J(w_i)$ for all $i \m$. More or less, most of algorithms take
advantage of this fact. The question is how to determine (all)
subsets $w_i \in \Wmin$ most effectively. At first, we illustrate
how the \KKA\ does it in the one-dimensional case.

\subsection{One-dimensional case}

As written above, for Example~\ref{example:p1} the set $\Wmin$
consists of $m = 10$ elements and the set $\Hset$ contains 9
points $h_1,\ldots,h_9$. For each point $h_k, k \in \hat{9}$ there
exist exactly two subsets $w_{k_1}$ and $w_{k_2}$ from $\Wmin$
such that $(h_k,w_{k_1}) \in Z$ and $(h_k,w_{k_2}) \in Z$. These
two subsets correspond to two parabolas whose intersection has a
coordinate $h_k$ and that can be easily determined for a given
$h_k$ using the following algorithm (which works for arbitrary
$p$).
\begin{program} \label{prog:relation_Z}
How to find all $w \in \Q$ such that $(\beta, w) \in Z$ for a
given $\beta \in \R^{p}$:
\begin{enumerate}
    \item For all $i \n$ evaluate squared residual $r_i^2(\beta)$ and order
    them.Define $i_k \in \{1,\ldots,n\}$ by $r_{i_k}^2(\beta) =
    \res{k}(\beta)$ for all $k = 1, \ldots, n$.
    \item If $\res{h}(\beta) < \res{h + 1}(\beta)$ then return the unique subset
    $w$ such that $(w^i = 1 \Leftrightarrow r_i^2(\beta) \leq
    \res{h}(\beta))$.\\
    If $\res{h}(\beta) = \res{h + 1}(\beta)$, let us suppose
    that
    \begin{multline*}
        r_{i_1}^2(\beta) \leq \cdots \leq r_{i_l}^2(\beta) <
        r_{i_{l+1}}^2(\beta) = \cdots = r_{i_h}^2(\beta) = \\
        = r_{i_{h +1}}^2(\beta) = \cdots = r_{i_{l + t}}^2(\beta) < r_{i_{l + t +
        1}}^2(\beta) \leq \cdots \leq r_{i_n}^2(\beta).
    \end{multline*}
    Then return all the subsets $w$such that
    $I_w$ (see \eqref{eq:I_w}) contains $l$ indices corresponding to $i_1, \ldots,
    i_l$ and arbitrary $(h - l)$-element subset of indices
    corresponding to $i_{l + 1}, \ldots, i_{l + t}$. Hence, there
    are $t \choose h - l$ subsets in relation $Z$ with $\beta$.
\end{enumerate}
\end{program}

In general, if Assumption~\ref{assump:hyperplane} is fulfilled in
the case of $p = 1$, that for every $w_i \in \Wmin$ there exists
at least one point $h \in \Hset$ such that $(w_i,h) \in Z$. Taking
this into account, we can state: if the set $\Hset = \{h_1,
\ldots, h_{m-1}\}$ is known, all the subsets $\Wmin$ can be
obtained by performing Program~\ref{prog:relation_Z} for each $h_k
\in \Hset$.

Only remaining task is how to determine the set $\Hset$. According
to its definition, a point $\beta$ is an element of $\Hset$ if and
only if $\res{h}(\beta) = \res{h+1}(\beta)$, i.e. this equality is
sufficient and necessary condition. Since the condition contains
ordered residuals, it can not be used directly -- at first we need
to find some candidates for which we will perform ordering of
residuals. These candidates can be determined by the following
necessary condition: if $\beta \in \Hset$, then there are two
distinct indices $i,j$ such that $r_i^2(\beta) = r_j^2(\beta)$.
Let us denote the set containing all $\beta \in \R^{1}$ satisfying
this necessary condition by $H$, i.e.
\begin{equation}\label{eq:H_set_pre_def}
    H = \{\beta \in \R^{1}\,|\, r_i^2(\beta) = r_j^2(\beta), i \neq
    j\},
\end{equation}
obviously $\#H \leq 2{n \choose p + 1} =  2{n \choose 2} = n(n-1)$
(note that the equation is quadratic, i.e., it has two solutions
$\frac{y_i - y_j}{x_i - x_j}$ and $\frac{y_i + y_j}{x_i + x_j}$ --
we still assume that Assumption~\ref{assump:hyperplane} is
fulfilled, hence $x_i \neq \pm x_j$).

Now we already have all necessary for proposing a~one-dimensional
version of \KKA.
\begin{program}\label{prog:KKA_onedim}
    \KKA\ in the case of $p = 1$.\\
    Denote the elements of the set $Q^{(n,2)}$ by $\{v_1, \ldots, v_{n \choose
    2}\}$.
    \begin{enumerate}
        \item Set $k = 1$ and $J_{\text{min}} = +\infty$.
        \item Denote the indices of two data points from subset $v_k$ by
        $i,j$.\\
        Save solutions of equation $r_i^2(\beta) = r_j^2(\beta)$
        as $\beta_1$ and $\beta_2$, i.e., $\beta_1 = \frac{y_i - y_j}{x_i -
        x_j}$ and $\beta_2 = \frac{y_i + y_j}{x_i + x_j}$.
        \item Evaluate and order residuals $r_l^2(\beta_1), l = 1, \ldots, n$.
        \item If $\res{h}(\beta_1) = \res{h+1}(\beta_1)$ (i.e., $\beta_1 \in
        \Hset$), find subsets $w^{(1)}, w^{(2)} \in \Q$ which are
        in relation~$Z$ with $\beta_1$ (use Program~\ref{prog:relation_Z}).
        \item If $J(w^{(1)}) < J_{\text{min}}$, put $J_{\text{min}} =
        J(w^{(1)})$ and $\beta_{\text{min}} = \beta_1$. \\
        If $J(w^{(2)}) < J_{\text{min}}$, put $J_{\text{min}} =
        J(w^{(2)})$ and $\beta_{\text{min}} = \beta_2$.
        \item If $\beta_1 \neq \beta_2$, repeat last two steps for
        $\beta_2$ (modify $J_{\text{min}}$ and $\beta_{\text{min}}$ accordingly).
        \item If $k < {n \choose 2}$, put $k = k + 1$ and go back
        to step~2.
        \item Return $\beta_{\text{min}}$ as the LTS estimate for
        Problem~\ref{problem:LTS}.
    \end{enumerate}
\end{program}
This algorithm works is finite if the $\Hset$ contains only a
finite number of points. Assumption~\ref{assump:hyperplane} is a
sufficient condition for this; it is not a necessary condition,
but still it is very weak and easily verifiable.

\subsection{Multidimensional case}

In the case of $p > 1$, the situation is more complicated. The
source of complication is the fact, that the set $\Hset$ contains
infinitely many points. In order to resolve this problem, we need
to find some finite subset of $\Hset$, let us denote it $\Hset_p$,
having the following property: for every $w \in \Wmin$ there
exists $\beta \in \Hset_p$ such that $(\beta,w) \in Z$.

Analogously to the case of $p = 1$, we will be looking for
candidates for being an element of $\Hset_p$ in the set $H$,
namely in some suitable finite subset $H_p$ for which $\Hset_p
\subset H_p \subset H$. In the case of $p = 1$, the equality of
the type $r_i^2(\beta) = r_j^2(\beta)$ can have at most two
solutions, in the case of $p > 1$, this equality only
``decrements'' the dimension by 1, i.e., the dimension of its
solution is $p - 1$. But we need the dimension to be zero, this
can be reached by considering $p$ ``independent'' equations of the
type $r_i^2(\beta) = r_j^2(\beta)$, in other words, a system of
$p$ equations with $p$ unknowns $\beta^1, \ldots, \beta^p$
\begin{equation}\label{eq:Hp_system_predef}
    \begin{split}
        r_{i_1}^2(\beta) &= r_{i_2}^2(\beta) \\
        \vdots \ \ & \ \ \ \ \ \ \vdots \\
        r_{i_p}^2(\beta) &= r_{i_{p+1}}^2(\beta),
    \end{split}
\end{equation}
where $i_1, \ldots, i_{p+1}$ corresponds to one of $(p+1)$-element
subsets of $\hat{n} = \{1, \ldots, n\}$, i.e., to one element of
$\Qp$. Unfortunately, as we will prove later on, the system
\eqref{eq:Hp_system_predef} is equivalent to $2^p$ linear systems
of $p$ equations. If all these systems are regular, then original
system \eqref{eq:Hp_system_predef} can have up to $2^p$ solutions.
Taking into account this number and the fact that there are ${n
\choose p + 1} = \#\Qp$ different systems of type
\eqref{eq:Hp_system_predef}, the set $H_p$, which is to be defined
as a set of solutions of all such systems, contains ${n \choose
p+1}2^p$ points from $\R{p}$.

In the next section, all the sets introduced above (sets $H$,
$H_p$ and $\Hset_p$) will be redefined precisely and their
mentioned (and some others) properties will be proved. In
particular, we will propose assumptions which allow us to prove
that the set $\Hset_p$ is a suitable set for \KKA.

\subsection{Set $\Hset_p$}\label{subsec:Hset}

The goal of this section is to find a set $\Hset_p$ for which it
holds that for every $w \in \Wmin$ there exists at least one
$\beta \in \Hset_p$ such that $(\beta,w) \in Z$. We will define it
as it was hinted above, therefor we will need to define the sets
$H$ and $H_p$ containing candidates for being elements of $\Hset$
and $\Hset_p$.

The set $H$ is to be defined as in \eqref{eq:H_set_pre_def}. The
quadratic equation of the type $r_i^2(\beta) = r_j^2(\beta), i,j
\n$ is equivalent to two linear ones
\begin{equation} \label{eq:two_linear_eq}
    \begin{split}
    (x_i + x_j)^T \beta & = y_i + y_j\\
    (x_i - x_j)^T \beta & = y_i - y_j,
    \end{split}
\end{equation}
each defining $p-1$-dimensional hyperplane.
\begin{definition}
Let us denote
\begin{eqnarray*}
  H^{(i,j,\pm)} &=& \{\beta \in \R{p}\,|\, r_i^2(\beta) = r_j^2(\beta)\} = \{\beta \in \R{p}\,|\, y_i - x_i^T \beta = \pm  (y_j - x_j^T \beta) \}\\
  H^{(i,j,-)} &=& \{\beta \in \R{p}\,|\, y_i - x_i^T \beta = (y_j - x_j^T \beta) \} = \{\beta \in \R{p}\,|\, y_i - y_j = (x_i^T - x_j^T)\beta \} \\
  H^{(i,j,+)} &=& \{\beta \in \R{p}\,|\, y_i - x_i^T \beta = - (y_j - x_j^T \beta) \} = \{\beta \in \R{p}\,|\, y_i + y_j = (x_j^T + x_j^T)\beta \}
\end{eqnarray*}
for every $i,j \n$, $i \neq j$ and
\begin{eqnarray*}
H &=& \cup_{i,j \n, i\neq j}H^{(i,j,\pm)} = \left( \cup_{i,j \n,
i\neq j}H^{(i,j,+)} \right) \cup \left( \cup_{i,j \n, i\neq
j}H^{(i,j,-)} \right).
\end{eqnarray*}
\end{definition}
Obviously the set $H$ from this definition is the same one as the
definition given in \eqref{eq:H_set_pre_def}.

It is apparent that $\Hset \subset H$. We also know, that the sets
$U_i, i \m$ from Definition~\ref{def:sequence_Ui} are separated by
the set~$\Hset$. The hyperplanes $H^{(i,j,+)}$ and $H^{(i,j,-)}$
divide $\R{p}$ into \emph{closed} convex sets, so-called polytops,
let us denote them $P_k$, $k \in \hat{K}$, where $K$ is some
finite number. We get that $\cup_{k \in \hat{K}}\partial P_k = H
\supset \Hset = \cup_{i \in \hat{m}}
\partial U_i$ and it implies that for every $U_i$ there exist
convex sets $P_{k_1}, \ldots P_{k_l}, l \geq 1$ such that
$\bar{U_i} = \cup_{j \in \hat{l}} P_{k_j}$ and also that $m \leq
K$.

It is well known fact, that a bounded polytop equals to a convex
envelope of points which are intersections of the hyperplanes
bordering the polytop. The smallest number of hyperplanes allowing
their intersection to be a point (i.e. the set with dimension 0)
equals the dimension of the space. For us the space is $\R{p}$ and
the dimension is $p$. We propose some more notation to express
what this fact means for our particular case.

Let $\circ \in \{+,-\}$ represent one of the arithmetical
operations, either an addition or a subtraction, i.e. $x \circ y =
x + y$ if $\circ = +$ and $x \circ y = x - y$ if $\circ = -$.

Let $\beta \in H$ be an intersection of $q + 1$, $q \geq 1$, sets
of the type $H^{(i,j,\pm)}$ such that $\beta \in H^{(i_1,i_2,\pm)}
\cup H^{(i_2,i_3,\pm)} \cup \cdots \cup H^{(i_q,i_{q+1},\pm)}$
where $i_1, \ldots , i_{q+1} \n$ are distinct. It means that
$\beta$ is a solution of the following system
\begin{equation*}
    \begin{split}
      r_{i_1}^2(\beta) &= r_{i_2}^2(\beta) \\
     \vdots \ \ & \ \ \ \ \ \ \vdots \\
      r_{i_q}^2(\beta) &= r_{i_{q+1}}^2(\beta).
     \end{split}
\end{equation*}
Note that if $r_{i_1}^2(\beta) = \res{h}(\beta) =
\res{h+1}(\beta)$ then, moreover, $\beta \in \Hset$. This system
of equations is equivalent to the following $2^q$ systems
\begin{equation} \label{eq:equations}
    \begin{split}
    (x_{i_1} \circ_1 x_{i_2})^T \beta & = y_{i_1} \circ_1
    y_{i_2}\\
    \vdots \qquad & \qquad  \vdots \\
    (x_{i_q} \circ_q x_{i_{q+1}})^T \beta & = y_{i_q} \circ_q
    y_{i_{q+1}}.
    \end{split}
    \end{equation}
where $(\circ_1, \dots, \circ_q)$ is an arbitrary element of the
set product $\times_{i = 1}^q \{+,-\}$.
\begin{definition} \label{def:set_Hset_p}
     Let $\beta \in H$ be a solution of system \eqref{eq:equations} of
     $q$ equations where $(\circ_1, \dots, \circ_q) \in \times_{i = 1}^q
     \{+,-\}$. Further, let us assume there exists no $i_{q+2}
     \n\setminus\{i_1,\ldots,i_{q+1}\}$ and $\circ_{q+1} \in \{+,-\}$ so that $(x_{i_{q+1}}
     \circ_{q+1} x_{i_{q+2}})^T \beta = y_{i_{q+1}} \circ_q y_{i_{q+2}}$ (i.e. $r^2_{i_1}(\beta) = \ldots = r_{i_{q+1}}^2(\beta) = r_{i_{q+2}}^2(\beta)$).
     Then we define an order of $\beta$
     \begin{equation*}
        \mathrm{Ord}(\beta) = \textrm{ number of linearly independent equations in \eqref{eq:equations}}.
     \end{equation*}
     We also define a set of \emph{zero-dimensional intersections}
     \begin{equation*}
        H_p = \{ \beta \in H\,|\,\mathrm{Ord}(\beta) = p \}
     \end{equation*}
     and its subset $\Hset_p$ of such $\beta \in H_p$ for which $r^2_{i_1}(\beta) = \ldots = r_{i_{p+1}}^2(\beta) =
     \res{h}(\beta) = \res{h+1}(\beta)$, where $i_1, \ldots, i_{p+1} \n$ are indices
     from the system of $p$ linearly independent equations of type \eqref{eq:equations}.
\end{definition}

Now we have all the notation necessary for proposing the assertion
which, despite being very simple and natural, will be used as the
basis of \KKA. At first, let us prove the following useful lemma
-- to be able to do it, we need this very weak assumption.
\begin{assump}\label{assump:hth_residual}
    \begin{equation*}
        (\forall \beta \in \R{p})(\res{h}(\beta) > 0),
    \end{equation*}
    i.e.,
    \begin{equation*}
        OF^{(LTS,n,h)}(\LTSbeta) > 0.
    \end{equation*}
\end{assump}
\begin{lemma}
    Let us assume that Assumptions \ref{assump:hyperplane} and \ref{assump:hth_residual} are fulfilled and let $B \subset \R{p}$ be a set containing all solutions of the
    system of equations
    \begin{equation} \label{eq:system_lemma}
    \begin{split}
    (x_{i_1} \circ_1 x_{i_2})^T \beta & = y_{i_1} \circ_1
    y_{i_2}\\
    \vdots \qquad & \qquad  \vdots \\
    (x_{i_q} \circ_q x_{i_{q+1}})^T \beta & = y_{i_q} \circ_q
    y_{i_{q+1}},
    \end{split}
    \end{equation}
    where $i_1, \ldots, i_{q+1} \n$ are distinct, $q \geq 1$, $\circ_1,\ldots,\circ_q \in
    \{+,-\}$ and $r_{i_1}^2(\beta) = \cdots = r_{i_{q+1}}^2(\beta)= \res{h}(\beta) = \res{h+1}(\beta)$. Then for all $j,k \in \widehat{q+1}, j \neq k$ and
    $\circ \in \{+,-\}$ either
    \begin{equation} \label{eq:res_equals_res_or_not}
    (\forall \beta \in B)((x_{i_j} \circ x_{i_k})^T \beta = y_{i_j} \circ
    y_{i_k})
    \end{equation}
    or
    \begin{equation*}
    (\forall \beta \in B)((x_{i_j} \circ x_{i_k})^T \beta \neq y_{i_j} \circ
    y_{i_k})
    \end{equation*}
    is true.

    Moreover, there exist $\circ_{l_1}, \ldots, \circ_{l_q} \in \{+,-\}$ such
    that system (\ref{eq:system_lemma}) is equivalent to the
    system
    \begin{equation} \label{eq:equations_alter}
    \begin{split}
    (x_{i_1} \circ_{l_1} x_{i_2})^T \beta & = y_{i_1} \circ_{l_1}
    y_{i_2}\\
    \vdots \qquad & \qquad  \vdots \\
    (x_{i_1} \circ_{l_q} x_{i_{q+1}})^T \beta & = y_{i_1} \circ_{l_q}
    y_{i_{q+1}}.
    \end{split}
    \end{equation}
    This holds even if $r_{i_1}^2(\beta) \neq \res{h}(\beta)$.
\end{lemma}

\proof \\
Assumption \ref{assump:hth_residual} implies that for all $s,t \n,
s \neq t$ the system
\begin{equation} \label{eq:lemma_pre_GAMFOG}
    \begin{split}
    (x_s + x_t)^T \beta & = y_s + y_t\\
    (x_s - x_t)^T \beta & = y_s - y_t
    \end{split}
\end{equation}
has not any solution $\beta \in \R{p}$ such that $r_s^2(\beta) =
r_t^2(\beta) = \res{h}(\beta)$. Indeed, if $\beta_0$ is a~solution
of both equations~\eqref{eq:lemma_pre_GAMFOG} and $r_s^2(\beta) =
r_t^2(\beta) = \res{h}(\beta)$,  then $r_s(\beta_0) =
r_t(\beta_0)$ and simultaneously $r_s(\beta_0) = - r_t(\beta_0)$
and so $\res{h}(\beta_0) = 0$, i.e. Assumption
\ref{assump:hth_residual} is not fulfilled (if $r_{(h)}(\beta_0) =
0$ then certainly $\beta_0 = \LTSbeta$).

Now, let us assume, without loss to generality, that $j < k$ holds
for $j,k$. Then for any $\beta \in B$
\begin{equation*}
    \begin{split}
    (x_{i_j} \circ_{j} x_{i_{j+1}})^T \beta & = y_{i_j} \circ_{j}
    y_{i_{j+1}}\\
    (x_{i_{j+1}} \circ_{j+1} x_{i_{j+2}})^T \beta & = y_{i_{j+1}} \circ_{j+1}
    y_{i_{j+2}}.
    \end{split}
\end{equation*}
If $j + 1 = k$, then, according to Assumption
\ref{assump:hth_residual} and the first paragraph of this proof,
equation \eqref{eq:res_equals_res_or_not} holds for all $\beta \in
B$ (when $\circ = \circ_{j}$) or it doesn't hold for any of them
(when $\circ \neq \circ_{j}$). Further, let $k$ be greater than $j
+ 1$. By adding together the equations, in the case of $\circ_j =
-$, or by subtracting them, in the case of $\circ_j = +$, we get
the equivalent system
\begin{equation*}
    \begin{split}
    (x_{i_j} \circ_{j} x_{i_{j+1}})^T \beta & = y_{i_j} \circ_{j}
    y_{i_{j+1}}\\
    (x_{i_j} \circ^{'}_{j+1} x_{i_{j+2}})^T \beta & = y_{i_{j}} \circ^{'}_{j+1}
    y_{i_{j+2}}\\
    (x_{i_{j+2}} \circ_{j+2} x_{i_{j+3}})^T \beta & = y_{i_{j+2}} \circ_{j+2}
    y_{i_{j+3}}.
    \end{split}
\end{equation*}
We can repeat this step $r$-times, $r \leq q$, until $j+r+1 = k$
and finish the proof of the first part of the lemma by employing
again assumption \ref{assump:hth_residual} for the system
\begin{equation*}
    \begin{split}
    (x_{i_j} \circ_{j} x_{i_{j+1}})^T \beta & = y_{i_j} \circ_{j}
    y_{i_{k}}\\
    (x_{i_j} \circ^{'}_{k-1} x_{i_{k}})^T \beta & = y_{i_{j}} \circ^{'}_{k-1}
    y_{i_{k}}.
    \end{split}
\end{equation*}

The second part of the lemma can be proved by repeating the same
steps for $j = 1$ and $k = q + 1$.

\proofend \\
The first part of the lemma tells us that each equation of the
type $(x_{i_j} \circ x_{i_{j+2}})^T \beta = y_{i_{j}} \circ
y_{i_{k}}, j,k \in \widehat{q + 1}, \circ \in \{+,-\}$ is either
equivalent with system \eqref{eq:system_lemma} or set of solutions
of this equation is disjoint with the set $B$ (due to
Assumption~\ref{assump:hth_residual}).

Now we can proceed to proving the most important assertion of this
section -- we shall prove that $\Hset_p$ is suitable (in a sense
of the previous section) set for \KKA. Is it true in general or we
have to assume that the data from \PLTS\ satisfies some condition?
The answer is that we have to propose some assumption which is,
however, quite weak. The reason is that it could happen that the
set $\Hset_p$ is empty. For example, in the case of $p = 2$, if
all hyperplanes of the type $H^{(i,j,\circ)}$ are parallel, then
there is no zero-dimensional intersection. It happens if and only
if all vectors $x_i$ from \PLTS\ are parallel. Then all systems of
equations of the type
\begin{equation*}
    \begin{split}
    (x_{i_1} \circ_{1} x_{1_2})^T \beta & = y_{i_1} \circ_{1}
    y_{i_{2}}\\
    (x_{i_1} \circ_{2} x_{i_3})^T \beta & = y_{i_{1}} \circ_{2}
    y_{i_{3}}
    \end{split}
\end{equation*}
are linearly dependent and so they have either no solution or the
set of all solutions is one-dimensional hyperplane. To avoid such
a situation we propose the following assumption.
\begin{assump}\label{assump:p_full_rank}
 For all $(\circ_1, \ldots, \circ_{n-1}) \in \times^n_{i = 1} \{+,-\}$ the matrix of
the system of $n-1$ equations
\begin{equation*}
    \begin{split}
    (x_{1} \circ_{1} x_2)^T \beta & = y_{1} \circ_{1}
    y_{2}\\
    \vdots \qquad & \qquad  \vdots \\
    (x_{1} \circ_{n-1} x_{n})^T \beta & = y_{1} \circ_{n-1}
    y_{n}
    \end{split}
\end{equation*}
has rank $p$.
\end{assump}
Obviously, it prevents the situation described above. This
assumption has a consequence which will be crucial for the proof
of the main assertion. Let us formulate it as a lemma.
\begin{lemma}\label{lemma:H_p_proof}
Let us assume that data from \PLTS\ satisfies
Assumption~\ref{assump:p_full_rank}. If $p
> 1$, then for
$$
(\forall l \in \{2,\ldots,p\})(\forall i_1,\ldots,i_l \n)(\forall
\circ_1, \ldots,\circ_{l-1} \in \{+,-\})
$$
it holds that if the system of $l - 1$ equations
\begin{equation} \label{eq:system_lemma_pre_KKA}
    \begin{split}
    (x_{i_1} \circ_{1} x_{i_2})^T \beta & = y_{i_1} \circ_{1}
    y_{i_2}\\
    \vdots \qquad & \qquad  \vdots \\
    (x_{i_1} \circ_{l-1} x_{i_l})^T \beta & = y_{i_1} \circ_{l-1}
    y_{i_l}
    \end{split}
\end{equation}
has rank $l-1$, then there exist $i_{l+1} \in \hat{n}\setminus
\{i_1,\ldots,i_l\}$ and $\circ_l \in \{+,-\}$ such that the system
of $l$ equations
\begin{equation*}
    \begin{split}
    (x_{i_1} \circ_{1} x_{i_2})^T \beta & = y_{i_1} \circ_{1} y_{i_2}\\
    \vdots \qquad & \qquad  \vdots \\
    (x_{i_1} \circ_{l-1} x_{i_l})^T \beta & = y_{i_1} \circ_{l-1} y_{i_l}\\
    (x_{i_1} \circ_{l} x_{i_{l+1}})^T \beta & = y_{i_1} \circ_{l} y_{i_{l+1}}
    \end{split}
\end{equation*}
has rank $l$.

Moreover, if $\beta \in \R{p}$ is a solution of
\eqref{eq:system_lemma_pre_KKA} such that $r_{i_1}^2(\beta) =
\res{h}(\beta) = \res{h+1}(\beta)$ then $i_{l+1}$ can be selected
so that there exists $\beta_1$ such that $r_{i_1}^2(\beta_1) =
\cdots = r_{i_{l}}^2(\beta_1) = r_{i_{l+1}}^2(\beta_1) =
\res{h}(\beta_1) = \res{h+1}(\beta)$
\end{lemma}

\proof

Let us denote the elements of the set of indices
$\hat{n}\setminus\{ i_1, \ldots, i_l \}$ by $\{j_{l+1}, \ldots,
j_{n}\}$ and let us select signs $\circ_{l}, \ldots, \circ_{n-1}$
arbitrarily. Then, due to Assumption~\ref{assump:p_full_rank}, the
system of equation
\begin{equation*}
    \begin{split}
    (x_{i_1} \circ_{1} x_{i_2})^T \beta & = y_{i_1} \circ_{1}
    y_{i_2}\\
    \vdots \qquad & \qquad  \vdots \\
    (x_{i_1} \circ_{l-1} x_{i_{l}})^T \beta & = y_{i_1} \circ_{l-1}
    y_{i_{l}}\\
    (x_{i_1} \circ_{l} x_{j_{l+1}})^T \beta & = y_{i_1} \circ_{1}
    y_{j_{l+1}}\\
    \vdots \qquad & \qquad  \vdots \\
    (x_{i_1} \circ_{n-1} x_{j_n})^T \beta & = y_{i_1} \circ_{n-1}
    y_{j_n}
    \end{split}
\end{equation*}
has rank $p$. We also know that $l-1$ vectors
$$
((x_{i_1} \circ_{1} x_{i_2}),\ldots,(x_{i_1} \circ_{1} x_{i_l}))
$$
are linearly independent. Then, due to Steinitz Theorem, there
exist indices $j_{l+1}, \ldots, j_{p+1}$ and $k_l, \ldots, k_p$
(in other words, there exist $p-l$ rows of the matrix of the
system above) such that $p$ vectors
$$
((x_{i_1} \circ_{1} x_{i_2}),\ldots,(x_{i_1} \circ_{l-1}
x_{i_l}),(x_{i_1} \circ_{k_l} x_{j_{l+1}}),\ldots,(x_{i_1}
\circ_{k_{l+1}} x_{j_p}))
$$
form a base of the vector space $\R{p}$. Obviously, as the index
$i_{l+1}$ can be then taken each index from $\{j_{l+1}, \ldots,
j_{p+1}\}$.

The second part of the lemma is a consequence of the first one and
the continuity of squared residuals. Let us denote the set of all
solutions of the system~\eqref{eq:system_lemma_pre_KKA} by $B
\subset \R{p}$. We know that there exist $\beta, \beta_1 \in B$
and $j_{l+1}$ such that $r_{i_1}^2(\beta) = \cdots =
r_{i_{l}}^2(\beta) = \res{h}(\beta)$ and $r_{i_1}^2(\beta_1) =
\cdots = r_{j_{l+1}}^2(\beta_1)$. Than, due to continuity of
squared residuals, there must exists $\beta_2 \in B$ and $i_{l+1}$
such that $r_{i_1}^2(\beta_2) = \cdots = r_{i_{l+1}}^2(\beta_2) =
\res{h}(\beta_2) = \res{h+1}(\beta_2)$. Note that it could happen
that $\beta_2 = \beta$ or $\beta_2 = \beta_1$.

\proofend

\begin{assertion} \label{asser:gamfog_base}
    Let us assume that Assumptions \ref{assump:hyperplane}, \ref{assump:hth_residual} and
    \ref{assump:p_full_rank} are fulfilled. If for the set $U_i \in \Useq, i \in \hat{m}$ holds that $\partial U_i \neq \emptyset$, then
    \begin{equation*}
        (\exists \beta \in \R{p}) (\beta \in \partial U_i \cap \Hset_p),
    \end{equation*}
    i.e. there exist $i_1, \ldots i_{p+1} \n$ and $\circ_1, \ldots, \circ_p \in \{+,-\}$
    such that $\beta$ is the only one solution of the system of $p$ linearly independent
    equations
    \begin{equation*}
    \begin{split}
    (x_{i_1} \circ_{1} x_{i_2})^T \beta & = y_{i_1} \circ_{1} y_{i_2}\\
    \vdots \qquad & \qquad  \vdots \\
    (x_{i_p} \circ_{p} x_{i_{p+1}})^T \beta & = y_{i_p} \circ_{p}
    y_{i_{p+1}},
    \end{split}
    \end{equation*}
    where moreover $r_{i_1}^2(\beta) = \res{h}(\beta) = \res{h+1}(\beta)$ is true.
\end{assertion}

\proof

Within the proof we shall partially use a syntax of computer
programming -- $q$ will be treated as a variable which is a
parameter of a loop, hence $q = q+1$ means incrementing $q$ of 1.
The instruction \emph{go to $\heartsuit$} means ``go back to the
line beginning with the sign $\heartsuit$''. Let us also suppose
that $p > 1$, if $p=1$, then the assertion is trivial.

Put $q = 1$. As $\partial U_i \neq \emptyset$, there are
$\beta^{(1)} \in \partial U_i$, $i_1 \in I_{w_i}$ and $i_2 \in
O_{w_i}$ (see~\eqref{eq:I_w} and for $w_i$ see
Definition~\ref{def:sequence_Ui}) so that $\res{h}(\beta^{(1)}) =
\res{h+1}(\beta^{(1)}) = r_{i_1}^2(\beta^{(1)}) =
r^2_{i_2}(\beta^{(1)})$, i.e. there exists $\circ_1$ such that
$\beta^{(1)} \in H^{(i_1,i_2,\circ_1)}$, i.e. $\beta^{(1)}$ is a
solution of equation
$$
(x_{i_1} \circ_{1} x_{i_2})^T \beta = y_{i_1} \circ_{1} y_{i_2}.
$$
According to the previous lemma, there exist $i_3 \n$, $\circ_2
\in \{+,-\}$ and $\beta^{(2)} \in \R{p}$ such that $\beta^{(2)}$
is a solution of the system
\begin{equation*}
    \begin{split}
    (x_{i_1} \circ_{1} x_{i_2})^T \beta & = y_{i_1} \circ_{1} y_{i_2}\\
    (x_{i_1} \circ_{2} x_{i_{3}})^T \beta & = y_{i_1} \circ_{2}
    y_{i_3},
    \end{split}
\end{equation*}
where moreover $r_{i_1}^2(\beta^{(2)}) = \res{h}(\beta^{(2)}) = \res{h+1}(\beta^{(2)})$.\\

$\heartsuit$ \ Put $q = q + 1$. If $q = p$, the proof is
completed. If it is not, the indices $i_1, \ldots, i_q$, signs
$\circ_1, \ldots, \circ_{q-1}$ satisfy the assumptions of the
previous lemma. Hence, there is an index $i_{q+1}$, a sign
$\circ_q$ and $\beta^{(q)}$ such that
\begin{equation*}
    \begin{split}
    (x_{i_1} \circ_{1} x_{i_2})^T \beta^{(q)} & = y_{i_1} \circ_{1} y_{i_2}\\
    \vdots \qquad & \qquad  \vdots \\
    (x_{i_q} \circ_{q} x_{i_{q+1}})^T \beta^{(q)} & = y_{i_q} \circ_{q}
    y_{i_{q+1}}
    \end{split}
\end{equation*}
and $r_{i_1}^2(\beta^{(q)}) = \res{h}(\beta^{(q)}) = \res{h+1}(\beta^{(q)})$ \\
Go to~$\heartsuit$. \proofend

The assertion ensures us that $\Hset_p$ is a suitable for \KKA\
and so we can propose also the multidimensional version of \KKA.
At the end of this section let us propose the following corollary.
\begin{corollary}\label{corollary:m_estimate}
Let us assume that Assumptions \ref{assump:hyperplane},
\ref{assump:hth_residual} and \ref{assump:p_full_rank} are
fulfilled. For the number $m = \#\Useq$ (see
Definition~\ref{def:sequence_Ui}) we have
\begin{equation*}
    m \leq {n \choose p+1}2^p.
\end{equation*}
\end{corollary}

\proof

    Due to Assertion~\ref{asser:gamfog_base} we know that $m \leq
    \#\Hset_p$. Then, the proof follows from the fact that $\Hset_p
    \subset H_p$, where $\# H_p \leq {n \choose p+1}2^p$.

\proofend

\section{\KKA-- new exact algorithm}

\subsection{Description of algorithm}

The multidimensional version of \KKA\ is a straightforward
generalization of Program~\ref{prog:KKA_onedim}. Note that if
Assumption~\ref{assump:hyperplane} is satisfied, then in the case
of $p = 1$ it holds that $\Hset = \Hset_p$. If $p > 1$ then
$\Hset_p \subsetneq \Hset$. In short, \KKA\ can be described as
follows: find all $\beta \in H_p$, by ordering the residuals
verify whether $\beta$ is also element of $\Hset_p$. If it is,
find all $w \in \Q$ such that $(\beta, w) \in Z$ and evaluate
$J(w)$ (see \eqref{eq:JW_def}) for them. The minimal obtained
value then corresponds to $w^*$ (see \eqref{eq:wstar_def}).

\begin{program}\label{prog:KKA}
    \KKA\ -- generic definition for all dimensions. \\
    Denote all the elements of the set $Q^{(n,p+1)}$ by $\{v_1, \ldots, v_{n \choose
    p+1}\}$ and all the elements
    of the cartesian product $\times_{j = 1}^p \{+,-\}$ by $\{\circ^{(1)}, \ldots, \circ^{(2^p)}\}$,
    where $\circ^{(i)} = (\circ^{(i)}_1, \ldots, \circ^{(i)}_p), i = 1, \ldots 2^p$.
    \begin{enumerate}
        \item Set $k = 1$ and $J_{\text{min}} = +\infty$.
        \item If $k > {n \choose p + 1}$ go to step~9.
        \item Denote the indices of data from subset $v_k$ by
        $i_1,\ldots, i_{p+1}$, hence $i_1,\ldots,i_{p+1} \n$ are distinct and $v^{i_1}_k = \cdots = v^{i_{p+1}}_k =
        1$. Put $l = 1$.
        \item If $l > 2^p$, put $k = k + 1$ and go to step~2.
        \item If the system of equations
            \begin{equation}\label{eq:equations_GAMFOG_descr}
                \begin{split}
                (x_{i_1} \circ^{(l)}_{1} x_{i_2})^T \beta & = y_{i_1} \circ^{(l)}_{1} y_{i_2}\\
                \vdots \qquad & \qquad  \vdots \\
                (x_{i_1} \circ^{(l)}_{p} x_{i_{p+1}})^T \beta & = y_{i_1} \circ^{(l)}_{p}
                y_{i_{p+1}}
                \end{split}
            \end{equation}
            is regular, then denote its solution by $\beta_0$, if
            it is not, put $l = l+1$ and go to step~4.
        \item Evaluate and order residuals $r^2(\beta_0)$.
        \item If $r_{i_1}^2(\beta_0) = \res{h}(\beta_0) = \res{h+1}(\beta_0)$,
        find subsets $w^{(1)}, \ldots, w^{(g)} \in \Q$ which are
        in relation~$Z$ with $\beta_0$ (use Program~\ref{prog:relation_Z}).
        \item For $j = 1, \ldots, g$ evaluate $J(w^{(j)})$.
        If $J(w^{(j)}) < J_{\text{min}}$, put $J_{\text{min}} =
        J(w^{(j)})$ and $w_{\text{min}} = w^{(j)}$.
        \item Put $l = l+1$ and go to step~4.
        \item Return $\beta = \hat{\beta}^{(OLS,W_{\text{min}}X,W_{\text{min}}Y)},
        W_{\text{min}} = \text{diag}(w_{\text{min}})$ as the LTS estimate for
        Problem~\ref{problem:LTS}.
    \end{enumerate}
\end{program}
Several of steps need to be commented on in details. Prior to
that, let us prove that \KKA\ always find the exact LTS estimate
for Problem~\ref{problem:LTS}.
\begin{theorem}
If data of Problem~\ref{problem:LTS} satisfy
Assumptions~\ref{assump:hyperplane}, \ref{assump:hth_residual}
and~\ref{assump:p_full_rank}, then \KKA\ returns the LTS
estimate~$\LTSbeta$.
\end{theorem}
\proof

The proof follows straightforwardly from
equation~\eqref{eq:OF_LTS_reformulation} and
Assertion~\ref{asser:gamfog_base}, which tells us that during
\KKA\
 we evaluate $J(w)$ for all $w \in \Wmin$.

 \proofend

Now, let us comment on the steps.
\begin{description}
    \item[steps 2 -- 5]
        The goal of the algorithm is to find all points of the set
        $\Hset_p$ and for every $\beta \in \Hset_p$ find all $U_i, i \in \hat{m}$
        and the corresponding $w_i$ such that $\beta \in \partial
        U_i$. In order to find all elements of $\Hset_p$, it is necessary to find all
        elements of $H_p$ and it requires to resolve all possible systems of equations of type
        \eqref{eq:equations_GAMFOG_descr}, i.e. it is necessary to go through all ${n \choose p+1}2^p$
        possibilities.
    \item[step 5] Due to assertion \ref{asser:gamfog_base} we know that ``in most cases'' we can omit non-regular systems
    without losing the assurance that we will find all the elements of $\Hset_p$ (see the
    proof). Non-regularity would become a problem only if Assumption~\ref{assump:p_full_rank}
    is disrupted, i.e. at least one matrix $(n-1 \times p+1)$ would not be regular. The
    assumptions will be discussed later on.
    \item[step 7] In this step $\beta_0$ is surely an element of
    $H_p$, to verify that $\beta_0 \in \Hset_p$ it is necessary
    to verify whether $r_{i_1}^2(\beta_0) = \res{h}(\beta_0) =
    \res{h+1}(\beta_0)$. If $r_{i_1}^2(\beta_0) \neq \res{h}(\beta_0) =
    \res{h+1}(\beta_0)$, then either $\beta_0 \notin \Hset_p$ or it
    will be found during the loop for another value of the parameter~$k$.

    If we assume that the equality $r^2_{i_1}(\beta_0) =
    \res{k}(\beta_0)$ has the same probability for all $k \n$ and for all $\beta_0 \in H_p$,
    then the probability that the algorithm will go to step 8 from step
    7 (i.e. that $r^2_{i_1}(\beta_0) =
    \res{h}(\beta_0) = \res{h+1}(\beta_0)$ holds) is $p/(n - p + 1)$.
    \item[step 7 and 8] How many $w$ can be in relation $Z$ with
    $\beta_0$? The number $g$ depends on $l \in \{0,1,\ldots,p-2\}$ for which
    $r_{i_1}^2(\beta_t) = \res{h - l}(\beta_t)$ and equals $p \choose l +
    1$. The worst case is ${p \choose [p/2]}$ and the best one is
    $p$ (see also Program~\ref{prog:relation_Z}).
\end{description}

To conclude the basic description of the algorithm, we will calculate the complexity of it. \\
In order to compute the exact LTS estimate $\LTSbeta$ by \KKA\ it
is necessary to
\begin{itemize}
    \item successively select all $n \choose p+1$ elements of the set
    $Q^{(n,p+1)}$,
    \item ${n \choose p+1}\cdot 2^p$ times resolve system
    \eqref{eq:equations_GAMFOG_descr} of $p$ equations,
    \item ${n \choose p+1}\cdot 2^p$ times evaluate and order $n$ residuals,
    \item ${n \choose p+1}\cdot 2^p \cdot \frac{p}{n - p + 1} \cdot {p \choose
    [p/2]}$ times calculate the OLS estimate $\OLSbetaW$.
\end{itemize}
Of course, the complexity further depends on numerical methods
used for ordering of the residuals and solving the systems of
equation (in step 4 and also during calculating the OLS estimate
in step 6) but such a discussion is beyond the scope of this work.

\subsection{Assumptions -- verification and disruption of them}

As written above, Assumption~\ref{assump:hyperplane} is quite weak
and moreover easily verifiable.
Assumption~\ref{assump:hth_residual} is still weaker and we can
rely on it without any doubt.

Concerning Assumption \ref{assump:p_full_rank} the situation is a
bit more complicated. To verify that all $2^{n-1}$ matrixes have
full rank is too exhausting. On the other hand, an assumption that
$(n-1 \times p)$ matrix has rank $p$ is quite weak (for $n$ great
enough) and we can rely on this is fulfilled.

However, if an intercept is considered,
Assumption~\ref{assump:p_full_rank} is always disrupted for
$\circ_1 = \cdots = \circ_{n-1} = -$ for the first column of the
matrix contains only zeros and so the rank of the matrix is less
or equal to $p-1$. Thus, if an intercept is considered,
Assertion~\ref{asser:gamfog_base} (namely
Lemma~\ref{lemma:H_p_proof}) is not proved and we lose the
certainity that \KKA\ always finds the exact LTS estimate. To
resolve this problem, Assumption~\ref{assump:p_full_rank} has to
be reformulated to the following form.
\begin{assump}\label{assump:p_full_rank_intercept}
 For all $(\circ_1, \ldots, \circ_{n-1}) \in \left(\times^n_{i = 1} \{+,-\}\right) \setminus \{(-,\ldots,-)\}$ the matrix of
the system of equations
\begin{equation*}
    \begin{split}
    (x_{1} \circ_{1} x_2)^T \beta & = y_{1} \circ_{1}
    y_{2}\\
    \vdots \qquad & \qquad  \vdots \\
    (x_{1} \circ_{n-1} x_{n})^T \beta & = y_{1} \circ_{n-1}
    y_{n}
    \end{split}
\end{equation*}
has rank $p$.
\end{assump}
Assuming that this Assumption~\ref{assump:p_full_rank_intercept}
is fulfilled instead of Assumption \ref{assump:p_full_rank} we can
reprove Lemma~\ref{lemma:H_p_proof} for models where an intercept
is considered as follows.

\proof

The only difference between the proofs is selecting of the signs
$\circ_l, \ldots, \circ_{n-1}$. In the original Lemma we can
select them arbitrarily, here, if
Assumption~\ref{assump:p_full_rank_intercept} is considered, we
demand $(\exists k \in \{l, \ldots, n-1\} (\circ_k \neq -)$. The
rest of proof is completely the same.

\proofend

\section*{Conclusion}

BSA proved to be quick enough to be usable for reasonably large
data. Of course probabilistic algorithms are faster and they have
found the exact solution of Problem~\ref{problem:LTS} as well in
all cases the author tested. BSA algorithm has been implemented in
MATLAB and in C++ (by Roman Kápl) and is available by email.

\section*{Acknowledgement}

TBA

\bibliographystyle{plain}
\bibliography{LTS}

\end{document}